\documentclass[manuscript, authorversion]{acmart}

\AtBeginDocument{%
  \providecommand\BibTeX{{%
    \normalfont B\kern-0.5em{\scshape i\kern-0.25em b}\kern-0.8em\TeX}}}

\copyrightyear{2022}
\acmYear{2022}
\setcopyright{rightsretained}
\acmConference[UbiComp/ISWC '22 Adjunct]{Proceedings of the 2022
ACM International Joint Conference on Pervasive and Ubiquitous
Computing}{September 11--15, 2022}{Cambridge, United Kingdom}
\acmBooktitle{Proceedings of the 2022 ACM International Joint
Conference on Pervasive and Ubiquitous Computing (UbiComp/ISWC '22
Adjunct), September 11--15, 2022, Cambridge, United
Kingdom}\acmDOI{10.1145/3544793.3560323}
\acmISBN{978-1-4503-9423-9/22/09}

\begin{document}

\title{TeleSHift: Telexisting TUI for Physical Collaboration \& Interaction}

\author{Andrew Chen}
\affiliation{%
  \institution{National Taiwan University of Science and Technology}
  \city{Taipei}
  \country{Taiwan}}

\author{Tzu-Ling Yang}
\affiliation{%
  \institution{National Taiwan University}
  \city{Taipei}
  \country{Taiwan}}

\author{Shu-Yan Cheng}
\affiliation{%
  \institution{National Taiwan University of Science and Technology}
  \city{Taipei}
  \country{Taiwan}}

\author{Po-Sheng Cheng}
\affiliation{%
  \institution{National Taiwan University}
  \city{Taipei}
  \country{Taiwan}}

\author{Tzu-Han Lin}
\affiliation{%
  \institution{National Taiwan University of Science and Technology}
  \city{Taipei}
  \country{Taiwan}}

\author{Kaiyuan Lin}
\affiliation{%
  \institution{National Taiwan Normal University}
  \city{Taipei}
  \country{Taiwan}}

\renewcommand{\shortauthors}{Chen, et al.}

\begin{abstract}
    Current workflow on co-editing and simultaneous presentation of 3-D shapes is confined to on-screen manipulation, which causes loss of perceived information when presenting perceptual concepts or complex shapes between members. Thus, we create TeleSHift, a 3-D tangible user interface (TUI) with a telexisting communication framework for group-based collaboration and demonstration. In this work, we present a larger-scaled proof-of-concept prototype providing hands-on operation for shape-based interactions including multi-sided collaboration and one-to-many presentation. In contrast to previous works, we further extend the use of TUIs to support cooperative tasks with telexistence while enabling the linkage of manipulable bits to provide a better user experience and interactivity.
\end{abstract}




\maketitle
\section{Introduction}
Tangible User Interfaces (TUIs) enable humans to perceive spatial information more intuitively. Ever since its introduction, its applications have extended from simple static model-presenting tasks to providing touch-based feedback with a degree of interactivity. Even so, we consider that its applications have not been thoroughly explored and see the potential of TUIs in supporting the collaboration of physical prototyping and dimensional presentation. 

When it comes to presenting 3-D shapes and structures with 2-D interfaces, especially during simultaneous editing, perceptual loss of accessed information would inevitably occur. The concept of telexistence is to make two separated objects in different physical spaces to sense co-presence via interaction. On top of that, we proposed a 3-D TUI to achieve more realistic feedback and thorough perception, while enhancing interaction experience on both dimensional information knowing and object-initiated interpersonal exchanges.

This work contributes by exploring the use of TUIs in physically manipulating and presenting 3-D shapes while also creating a functional, larger-scaled proof-of-concept prototype with telexisting abilities to provide parallel display and simultaneous altering of shape-based information for multi-sided physical collaboration.

\section{Related Works}
Our work focuses on demonstrating cooperation and mutual knowing of dimensional modeling in different physical spaces with a programmable, shape-shifting TUI. Aside from exploring digital telexisting\cite{Tachi09} frameworks for physical-to-physical interactions\cite{Philippe14, Nishida17}, we focus on providing a better experience in physical modeling.

\textbf{Bit-Scaled Operations in Tangible Interfaces: }
The concept of \textit{Tangible Bits}\cite{Ishii1997} was incorporated to create more immersive interfaces with controllable small scaled bits to create a larger experience. Approaches to this concept include \textit{Materiable}\cite{Nakagaki16}, rendering material properties with bit-sized manipulation via physical simulation with a large tabletop interface. Works in this scope focus solely on the interactive interpretation of intended information, and this work aims to utilize such techniques into supporting collaborative physical modeling tasks.

\textbf{Interactivity in Physical Modelling: }
We see a gap within the scope of previous works regarding hands-on control of dimensions as manipulable area and methods of physical design is being limited, while only a handful of designated usages could be performed with its confined interaction methods, such as \textit{NURBSforms}\cite{Tahouni20} enables prototyping of curved surfaces with a magnetically actuated interface with the ability to record and load previously saved dimensions. Thus, in this work, we focus on exploring hands-on prototyping or presenting hardware for any 3-D shape on any scale.

This work aims to initiate an approach to extending such concepts into cooperation tasks, creating controllable self-deforming objects with the ability to connect with each other for a larger scale of interactivity while implementing a telexistence framework for collaboration in separate physical spaces.

\section{Implementation}
\begin{figure}
    \centering
    \includegraphics[width=\linewidth]{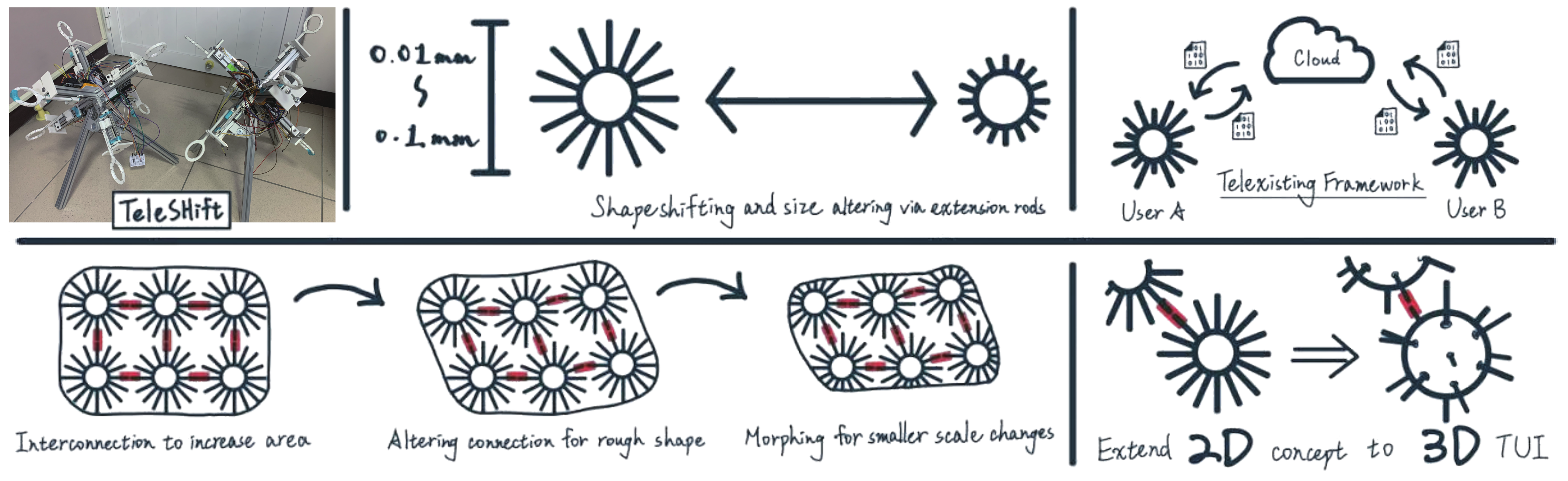}
    \caption{The storyboard illustrating the concept of TeleSHift along with our proof-of-concept prototype presented on the top left.}
    \label{fig:Storyboard}
\end{figure}

We present \textbf{TeleSHift}, which is a structure created by \textbf{Tele}xisting \textbf{sh}ape-\textbf{shift}ing substructures and used as a TUI. Our implementation of the substructure is constructed to have six extension arms able to perform 1-dimensional extend-interact movements. The arms, which consist of motorized slide potentiometers fixed onto the aluminum ridges, are each separated by 90 degrees to present deformation on the positive and negative of the x- y- and z-axes. Furthermore, the extending arms are also used to connect with other substructures by magnets for swarming to shape-shift on a larger scale, creating TeleSHifts.
The substructures’ control and communication are propelled with ESP32 and can exchange data with Google Realtime Firebase to update its current status including dimensions and jointing states, or fetch data to enable shape recovery and achieve synchronization with other connected TeleSHifts in different physical spaces.

The TeleSHift storyboard along with its larger-scaled proof-of-concept hardware interpretation is presented in Fig. ~\ref{fig:Storyboard}.

\section{Proposing Scenarios}
\subsection{Into Prototype Collaboration}
TeleSHift enables physical collaboration on editing the same 3-D structure in separate spaces effectively, breaking space restrictions and creating real-time interaction for consensus reaching on physical dimensions: When a group of designers physically in different spaces are working together on the creation of an ergonomic mouse, a common workflow would rely on 2-D interfaces of computer-aided design tools. However, a loss of perceived information, especially when collaborating simultaneously, would occur when presenting perceptual concepts or complex shapes between members. With TeleSHift, the designers can now work on the mouse’s prototype at the same time: Not only could they present the deformation of the created shape, but parallel knowing and simultaneous shape-shifting can also be achieved when one manually modifies their TeleSHift by presenting the same deformation status on the others simultaneously, and vice versa. 

\subsection{Into Dimensional Teaching}
TeleSHift also provides specific functionalities such as single-sided presentation, one-to-many demonstration, or saving shape-shift dimensions. An instance needing such requirements would be when a teacher wants to teach about the basic structure of a chair to several students and asks them to create their own design with some high-cost material.

The teacher can manually demonstrate the shaping process with TeleSHift, while students can watch and learn about the structural change with their own TeleSHifts at the same time. Moreover, students’ manipulation of their devices would not affect both the other students and the teacher, which allows students to come up with their creations. On the other hand, students can store the deformation status during their design at any time. When an undo process is required, they can return to the specific shape and redesign the chair. With TeleSHift as a substitute for expensive or raw materials, students can constantly reshape their objects, thus reducing cost and mitigating the waste of materials.

\section{Future Work}
The concept of TeleSHift is to enable two-staged shape-shifting for the presentation of any physical shapes at the users' discretion. This is achieved by substructure-to-substructure connection positioning for larger scale changes and self-deformation of the substructures to create smaller scale alterations.

In the future, we would like to work on scaling down the prototype and mass producing the substructures to take TeleSHift to its full potential. Also, as our current design only incorporates 6 extending arms to represent the axes in 3-D space, to add on more such structures would be worked on to provide a higher resolution of shape presentation. 

\begin{acks}
The authors would like to thank Yu-Ju Yang for valuable suggestions and assistance on scenario presenting as well as video making.
\end{acks}
\bibliographystyle{ACM-Reference-Format}
\bibliography{TeleSHift}

\end{document}